\documentclass[%
 reprint,
superscriptaddress,
 amsmath,amssymb,
 aps,
prl,
]{revtex4-1}

\usepackage{graphicx}
\usepackage{dcolumn}
\usepackage{bm}

\begin{document}


\title{Magic Sizes of Cationic and Protonated Argon Clusters}

\author{Michael Gatchell}
\email{michael.gatchell@uibk.ac.at}

\affiliation{Institut f\"{u}r Ionenphysik und Angewandte Physik, Universit\"{a}t Innsbruck, Technikerstr.~25, A-6020 Innsbruck, Austria}%
\affiliation{Department of Physics, Stockholm University, 106 91 Stockholm, Sweden}%

\author{Paul Martini}
\author{Lorenz Kranabetter}%
\affiliation{Institut f\"{u}r Ionenphysik und Angewandte Physik, Universit\"{a}t Innsbruck, Technikerstr.~25, A-6020 Innsbruck, Austria}%

\author{Bilal Rasul}%
\affiliation{Institut f\"{u}r Ionenphysik und Angewandte Physik, Universit\"{a}t Innsbruck, Technikerstr.~25, A-6020 Innsbruck, Austria}%
\affiliation{Department of Physics, University of Sargodha, 40100 Sargodha, Pakistan}

\author{Paul Scheier}
\affiliation{Institut f\"{u}r Ionenphysik und Angewandte Physik, Universit\"{a}t Innsbruck, Technikerstr.~25, A-6020 Innsbruck, Austria}%

\date{\today}

\begin{abstract}
There has long been a discrepancy between the size distributions of Ar$_n^+$ clusters measured by different groups regarding whether or not magic numbers appear at sizes corresponding to the closure of icosahedral (sub-)shells. We show that the previously observed magic cluster size distributions are likely the result of an unresolved Ar$_n$H$^+$ component, that is, from protonated argon clusters. We find that the proton impurity gives cluster geometries that are much closer to those for neutral rare gas clusters, which are known to form icosahedral structures, than the pure cationic clusters, explaining why the mass spectra from protonated argon clusters better matches these structural models. Our results thus show that even small impurities, e.g.\ a single proton, can significantly influence the properties of clusters. 

\end{abstract}

\maketitle


Rare gas clusters are some of the simplest chemical systems studied, with many of their structural properties deduced from basic sphere packing models \cite{Echt:1981aa} or classical two-body interactions such as the Lennard-Jones 6-12 potential \cite{Ikeshoji:1996aa,Rey:1992aa}. Theoretical studies have shown that the global energy minima of such clusters containing less less than a few hundred particles (with few exceptions \cite{Leary:1999aa}) prefer icosahedral geometries where shell closures (and the filling of faces on the polyhedra) are associated with enhanced stabilities \cite{Wales:1997aa,Xiang:2004ab,Xiang:2004aa}.


It is, alas, difficult to experimentally study the structures and stabilities of \emph{neutral} rare-gas clusters \cite{Bruhl:2004aa}. \emph{Charged} clusters are, however, easily studied using mass spectrometric techniques. In 1981, Echt \emph{et al.}\ reported that clusters of xenon formed by supersonic expansion and ionized by electron impact showed enhanced abundances at clusters sizes of 13, 19, 23, 25, 55, 71, 81, 87, 101, 135, and 147, which could be explained by sphere packing in the formation of icosahedral structures \cite{Echt:1981aa}. This was followed by numerous studies showing similar magic cluster size series in He$_n^+$ \cite{Stephens:1983aa,Buchenau:1990aa}, Ar$_n^+$ \cite{Harris:1984aa,Harris:1986aa}, Ne$_n^+$ \cite{Mark:1987aa}, and Kr$_n^+$ \cite{Lezius:1989aa,Miehle:1989ab} clusters. 

Early on it was noted that these series of magic cluster sizes were not always reproducible. This is particularly true in the case of Ar$_n^+$, where experimental results can be placed in one of two categories: 1) where increased abundances of Ar$_n^+$ clusters with $n = 13$, 19, 23, 26, 29, 32, 34, 43, 46, 49, 55, 64, 71, 81, and 87 are observed \cite{Harris:1984aa,Harris:1986aa}, and 2) where the most strongest anomaly of Ar$_n^+$ cluster sizes below $n=81$ is a particularly low abundance of Ar$_{20}^+$ \cite{Milne:1967aa,Ding:1983aa,Scheier:1987aa,Levinger:1988aa,Miehle:1989ab,Ferreira-da-Silva:2009aa}, which is considered to be anti-magic. The reasons for the observed differences have been debated for more than 30 years, mainly revolving around how the clusters are formed (e.g.\ whether the clusters are born neutral before being ionized, or are grown around a charged core), and remains poorly understood \cite{Vafayi:2015aa}.

In this paper we present high resolution mass spectrometry measurements of pure Ar$_n^+$ clusters and protonated Ar$_n$H$^+$ clusters that can be separated up to cluster sizes of $n\approx 100$. We find that the pure Ar$_n^+$ cluster series show few abundance anomalies in agreement with results from Refs.\ \cite{Milne:1967aa,Ding:1983aa,Scheier:1987aa,Levinger:1988aa,Miehle:1989ab,Ferreira-da-Silva:2009aa}, while the protonated Ar$_n$H$^+$ clusters show pronounced magic numbers in agreement with the results on Ar$_n^+$ clusters by Harris \emph{et al.}\ \cite{Harris:1984aa,Harris:1986aa} and neutral Lennard-Jones clusters \cite{Wales:1997aa}. We thus come to believe that the significant differences observed in past studies of argon clusters is due to contributions from protonated clusters. This conclusion is further motivated by \emph{ab initio} calculations of pure and protonated argon clusters.

We produce the argon clusters in He nanodroplets using the setup described in detail in Refs.\ \cite{Schobel:2011aa,Kurzthaler:2016aa,Kuhn:2016aa}. Briefly, droplets of He containing on average a few million atoms are formed by the supersonic expansion of compressed (2.5\,MPa) He through a nozzle that is cooled to 9.5\,K. The droplets capture Ar and H$_2$ gas, that is injected in a pickup chamber, which condense into clusters in the superfluid 0.37\,K droplets. The droplets are ionized by impact of 76\,eV electrons and the positively charged products are analyzed with a reflectron time-of-flight mass spectrometer (Tofwerk AG model HTOF). The mass spectra are calibrated and analyzed using the IsotopeFit software \cite{Ralser:2015aa}. This method of producing rare gas clusters has been used in the past to study Ar$_n^+$ \cite{Ferreira-da-Silva:2009aa} and Kr$_n^+$ \cite{Schobel:2011ab} clusters, giving results is good agreement with other techniques \cite{Milne:1967aa,Ding:1983aa,Scheier:1987aa,Levinger:1988aa,Miehle:1989ab}. 

\begin{figure*}[]
\includegraphics[width=7in]{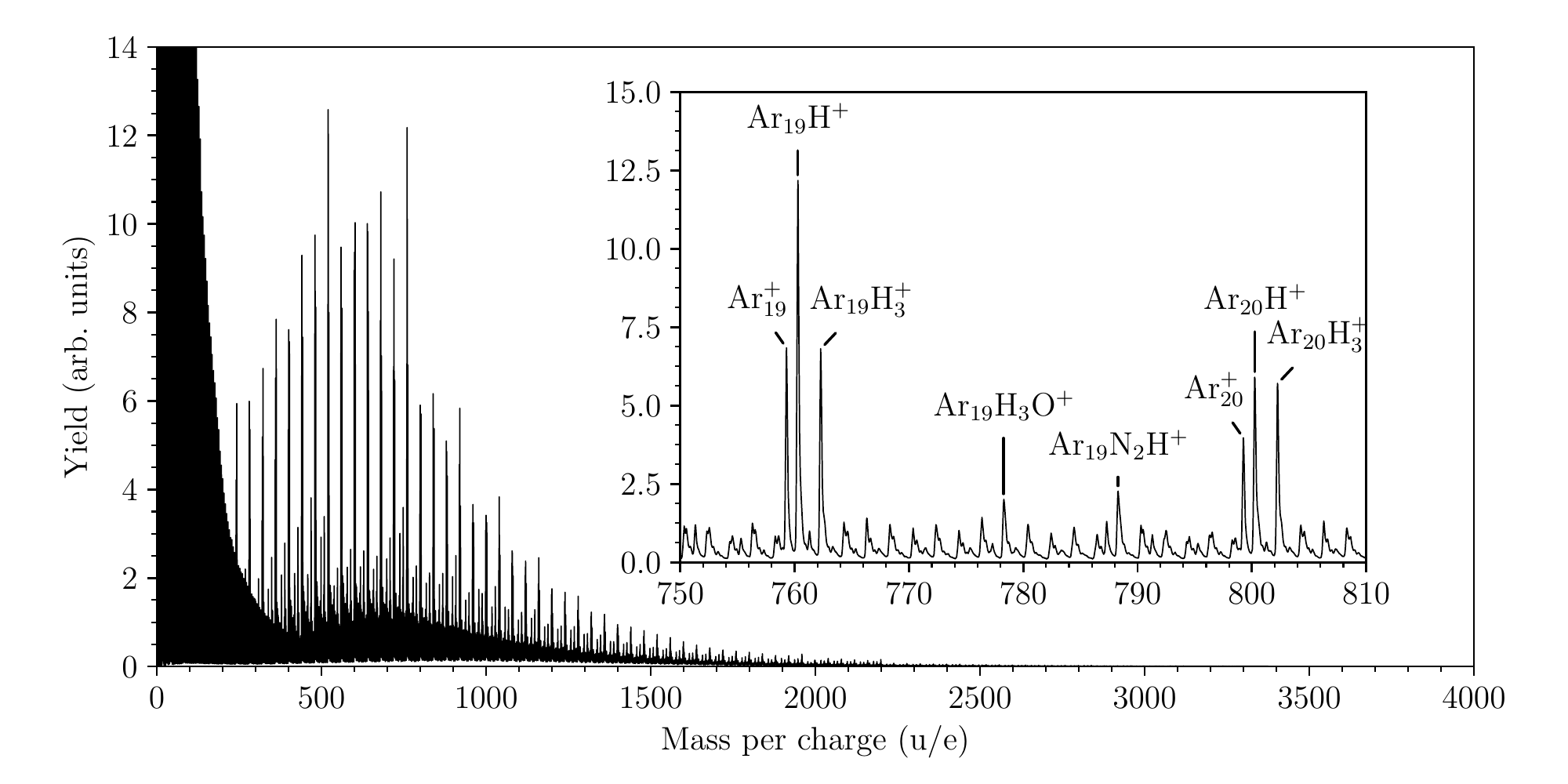}\hspace{2pc}%
\caption{Mass spectrum of positively charged products from He nandroplets doped with Ar and H$_2$ that are ionized by 76\,eV electrons. The distribution at low masses is the pure He$_n^+$ series and at higher masses clusters of Ar$_n^+$, Ar$_n$H$^+$, and Ar$_n$H$_3^+$ dominate. The inset shows a zoomed-in region of the spectrum where the particularly strong Ar$_{19}$H$^+$ peak is apparent.}
\label{fig:MS}
\end{figure*}

In Figure \ref{fig:MS} we show an overview spectrum from He nanodroplets doped with Ar and H$_2$ prior to ionization. At low masses (below about 400\,u/e) we mainly see the contribution from pure He$_n^+$ clusters from larger neutral droplets that fragment upon ionization. At higher masses the spectrum is dominated by Ar$_n^+$, Ar$_n$H$^+$, and Ar$_n$H$_3^+$ clusters that are free of helium. The inset of Figure \ref{fig:MS} shows a zoom-in of the mass range covering Ar$_{19}$X$^+$ and Ar$_{20}$X$^+$ systems. With the resolution of the mass spectrometer ($R\approx 4000$) we can clearly separate the individual cluster series, the relative intensities of which can be tuned by varying the Ar and H$_2$ pressures in the pickup region.
 
\begin{figure}[]
\includegraphics[width=3.5in]{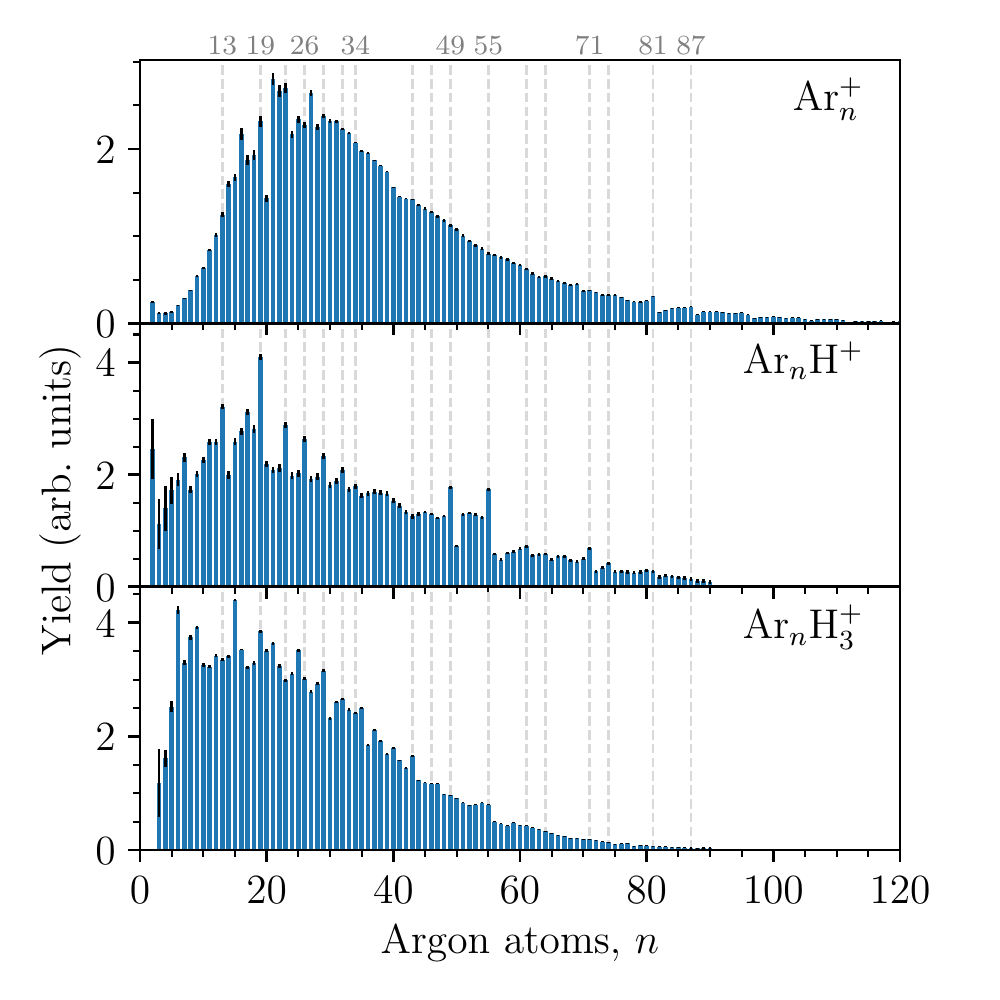}
\caption{Cluster size distributions of Ar$_n^+$, Ar$_n$H$^+$, and Ar$_n$H$_3^+$. Dashed lines show the magic cluster sizes reported by Harris \emph{et al.}\ \cite{Harris:1984aa,Harris:1986aa} for Ar$_n^+$. The statistical uncertainties are indicated by black errorbars.}
\label{fig:dists}
\end{figure}

Size distributions of Ar$_n^+$, Ar$_n$H$^+$, and Ar$_n$H$_3^+$ clusters are shown in Figure \ref{fig:dists} from separate measurements where we have optimized the intensities of each series. This does not affect the specific structures in each series, but can shift the underlying log-normal distributions that result from the pickup statistics. Vertical dashed lines in each panel show the magic cluster sizes reported by Harris \emph{et al.}\ \cite{Harris:1984aa,Harris:1986aa} for pure Ar$_n^+$ clusters and the most prominent of these features are labeled above the top panel. The pure Ar$_n^+$ series (top panel of Figure \ref{fig:dists}) displays a log-normal size distribution with few anomalies. The depleted Ar$_{20}^+$ channel reported numerous times in the past \cite{Milne:1967aa,Ding:1983aa,Scheier:1987aa,Levinger:1988aa,Miehle:1989ab,Ferreira-da-Silva:2009aa} is clearly visible, as are a few other anomalies. There is a clear drop-off in intensity between Ar$_{23}^+$ and Ar$_{24}^+$, one that has been reported before \cite{Ferreira-da-Silva:2009aa}, as well as what could be interpreted as magic peaks from Ar$_{16}^+$ and Ar$_{27}^+$. Although small on an absolute scale due to the underlying cluster size distirbution, there are also clear abundance anomalies visible for Ar$_{81}^+$ and Ar$_{87}^+$, which match magic numbers previously reported for Ar$_n^+$ \cite{Harris:1984aa,Harris:1986aa}, Kr$_n^+$ \cite{Lezius:1989aa}, and Xe$_n^+$ \cite{Echt:1981aa} clusters.

The distribution of protonated Ar clusters (middle panel of Figure \ref{fig:dists}) is clearly different from that of the pure argon clusters (top panel). It is immediately clear that every single magic size identified by Harris \emph{et al.}\ \cite{Harris:1984aa} for pure Ar$_n^+$ clusters is associated with an abundance anomaly in our Ar$_n$H$^+$ series. In addition to this, there are several more subtle features that agree between the two works, such as the particularly low abundance of Ar$_{50}$H$^+$ clusters that is followed by a plateau of relatively abundant Ar$_{51}$H$^+$ through Ar$_{54}$H$^+$ peaks. The main standout feature is that we also identify a magic Ar$_7$H$^+$ peak, which lies below the lower limit of most Ar$_n^+$ mass spectra found in the literature and is rarely discussed.

In the bottom panel Figure \ref{fig:dists} we show a size distribution of Ar$_n$H$_3^+$ clusters. Some anomalies match the magic sizes seen with Ar$_n$H$^+$ clusters (e.g.\ $n=19, 29, 43, 55$) though most do not. It is thus clear that the specific positions of abundance anomalies, i.e.\ magic numbers, is indeed dependent on the types of impurities present in the argon clusters. For the remainder of this letter we will mainly focus on the Ar$_n^+$ and Ar$_n$H$^+$ clusters.
 
Clusters of the form Ar$_n$X$^+$, where X is some impurity atom or molecule, have been studied in the past \cite{Hvistendahl:1990aa,Lezius:1992aa,Giju:2002aa,Ferreira-da-Silva:2009aa,McDonald:2016aa}, displaying magic features similar to those reported for pure Ar$_n^+$ clusters. For example, a magic Ar$_{54}$N$_2^+$ cluster has been identified where one of the Ar atoms in an icosahedral geometry is replaced with a N$_2$ molecule \cite{Ferreira-da-Silva:2009aa}. Protonated argon has also been well studied for small systems such as the ArH$^+$ dimer \cite{Bondybey:1972aa,Johns:1984aa} and the linear ArHAr$^+$ system \cite{Bondybey:1972aa,Kunttu:1994aa}. However, previous experimental studies on protonated argon clusters have only investigated small clusters containing less than 10 Ar atoms \cite{Hvistendahl:1990aa,McDonald:2016aa}, while theory has covered Ar$_n$H$^+$ sizes up to $n=35$ \cite{Giju:2002aa,Ritschel:2005aa,McDonald:2016aa}.

\begin{figure}[]
\includegraphics[width=3.5in]{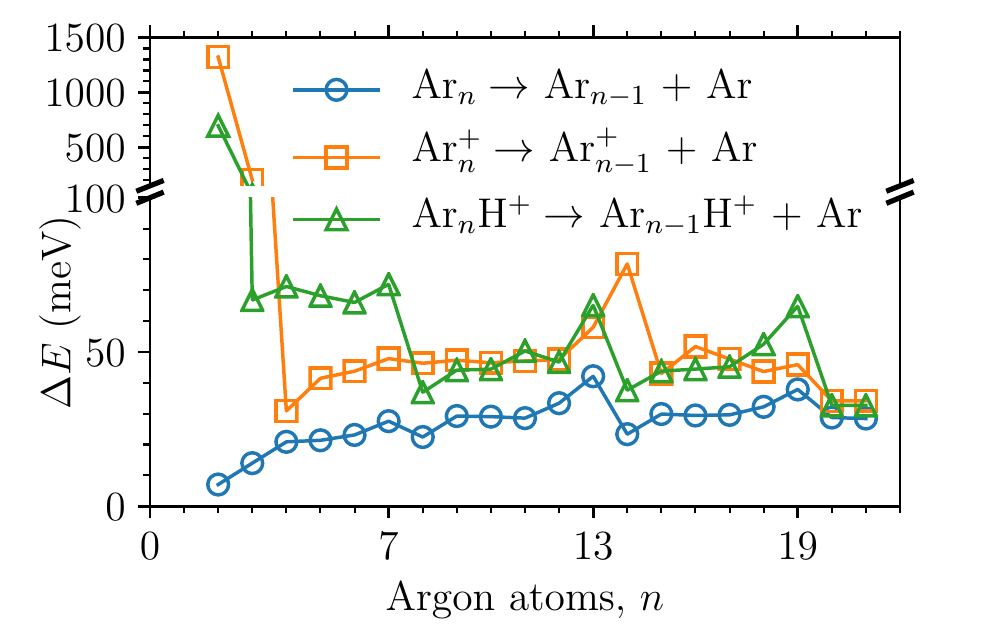}
\caption{Dissociation energy for losing a single Ar atom from different neutral and cationic clusters calculated at MP2(Full)/def2-SVPP level. The energies included zero-point corrections but not basis-set superposition error corrections.}
\label{fig:BEs}
\end{figure}

\begin{figure}[]
\includegraphics[width=3.5in]{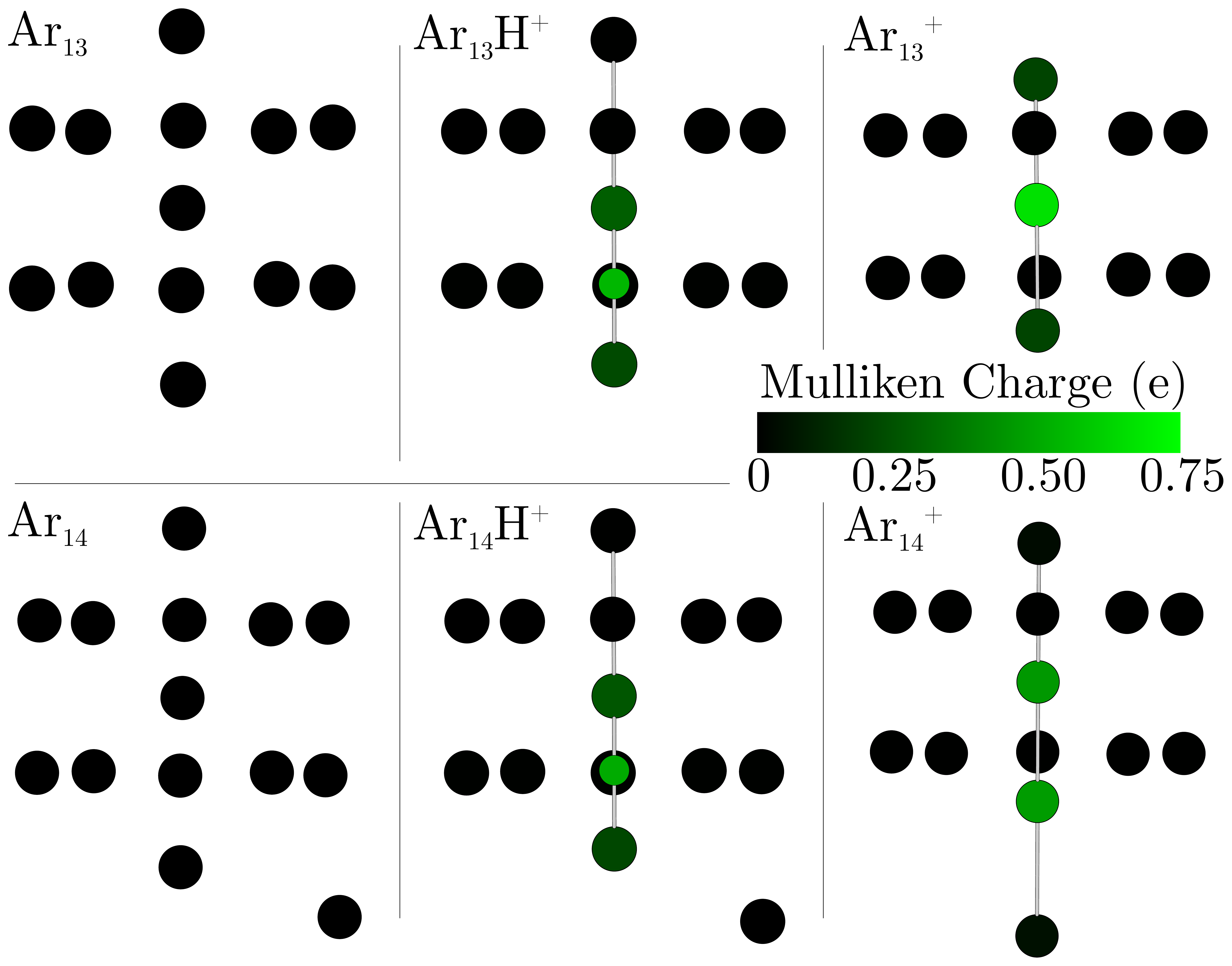}\\
\caption{Two-dimensional projections of icosahedral structures of Ar$_{13}$, Ar$_{13}$H$^+$, and Ar$_{13}^+$, and structures of Ar$_{14}$, Ar$_{14}$H$^+$, and Ar$_{14}^+$ optimized at MP2(Full)/def2-SVPP level. Atoms are colored based on their Mulliken charge and the proton in the center column is identified by a circle with a smaller radius that the rest. Coordinates of all calculated cluster geometries are given in the supplementary information.}
\label{fig:strucs}
\end{figure}

To better understand our experimental results, we have performed \emph{ab initio} structure calculations of neutral Ar$_n$ clusters, cationic Ar$_n^+$ clusters, and protonated Ar$_n$H$^+$ clusters for sizes up to $n=21$. The calculations were performed at MP2(Full)/def2-SVPP level using Gaussian 16 \cite{Frisch:2016aa_short}. This method was selected based on previous theoretical studies \cite{Giju:2002aa,McDonald:2016aa}, test-calculations on the geometries of small cluster sizes, and due to the favorable scaling that allows us to study clusters with up to relatively large sizes. The evaporation energy for losing a single, neutral Ar atom as a function of cluster size of these systems is shown in Figure \ref{fig:BEs} and the optimized structures of Ar$_{13}$, Ar$_{13}$H$^+$, Ar$_{13}^+$, Ar$_{14}$, Ar$_{14}$H$^+$, and Ar$_{14}^+$ in Figure \ref{fig:strucs}. The geometry optimizations were carried out starting from the structures of neutral Lennard-Jones clusters \cite{Wales:1997aa} and in the case of the pure Ar$_n^+$ clusters, geometries from Ref.\ \cite{Ikegami:1993aa} were also tested. The atomic coordinates for the lowest energy structure of each cluster is given in the supplementary information. 

In Figure \ref{fig:BEs} we can see that the curves for Ar$_n$ clusters (blue circles) and Ar$_n$H$^+$ clusters (green triangles) show the same main features, i.e.\ relatively tightly bound systems with $n=7, 13, 19$ followed by weaker systems for $n=8, 14, 20$. The main difference is that the protonated clusters are more tightly bound due the presence of the charge that attracts the surrounding argon atoms to the ArHAr$^+$ unit that forms the core of the cluster (see Figure \ref{fig:strucs}). Our calculations agree well with the structure determined in previous theoretical studies of protonated \cite{Giju:2002aa,Ritschel:2005aa,McDonald:2016aa} and cationic \cite{Ikegami:1993aa} argon clusters where sizes overlap and they readily explain the first few magic numbers observed in out experiments.

The purely cationic clusters (orange squares in Figure \ref{fig:BEs}) on the other hand show a very different behavior. We find no step in binding energy after $n=7$ and the first local maxima is instead located at $n=14$. The reason for the difference in this curve compared to the other two is due to the existence of an Ar$_3^+$ or Ar$_4^+$ core that these clusters form around. The linear Ar$_3^+$ core in Ar$_{13}^+$ is significantly contracted compared to other Ar--Ar distances as seen in Figure \ref{fig:strucs}. This strains the icosahedral geometry so that when a fourteenth Ar atom is added it interacts relatively strongly with this core, forming the basis for the Ar$_4^+$ system that is present in larger cluster sizes \cite{Ikegami:1993aa}. This elongated charged core means that there are more possible positions that other the Ar atoms in the cluster may interact with the charge-center, leading to competition between different cluster geometries so that there is less preference for a single dominant structure at small sizes (this could play a role for sizes smaller than 14 as well). This could explain the poorer agreement between the theory and experiments for the Ar$_n$ cluster distribution compared to the other systems. The optimal geometries of the protonated clusters on the other hand are very similar to the neutral systems, with the proton slotting in between two Ar atoms without significantly altering the distance between them (see Figure \ref{fig:strucs}). These compact structures are well explained by icosahedral geometries, giving the magic number series that is observed in the experiments.

The fact that the protonated argon clusters essentially retain the geometries of the neutral clusters is why the magic numbers predicted by sphere packing models are so well reproduced in the Ar$_n$H$^+$ series. The pure Ar$_n^+$ clusters instead behave as packed spheres with a structural defect (e.g.\ Ar$_3^+$) at their core \cite{Levinger:1988aa}. However, as the cluster sizes increase the overall effect of this distortion on the entire cluster will decrease. This likely the reason why abundance anomalies matching sphere packing models \cite{Echt:1981aa} begin to appear in the Ar$_n^+$ mass spectrum (Figure \ref{fig:dists}) for $n\geq 81$.

In light of the present results we have re-evaluated data from previous studies of Kr$_n^+$ clusters \cite{Schobel:2011ab} performed with the same setup as the current work. While that study did identify magic cluster sizes \cite{Schobel:2011ab}, we do not find any evidence that protonated clusters played a roll in those results. We suspect this is because for heavier rare gas clusters, the effect the charge has on the core of the clusters decreases, thus putting less strain on icosahedral packing of atoms. We thus do not believe that protonation plays an important role in the the magic series of Kr$_n^+$ \cite{Lezius:1989aa,Miehle:1989ab} and Xe$_n^+$ \cite{Echt:1981aa} clusters. It does, however, seem likely that for the lighter rare gases (Ne and He) protonation can have a strong effect on the geometries of charged clusters. Test calculations that we have performed on Ne$_n$H$^+$ clusters show a similar behavior as we see for argon clusters, i.e.\ that protonated clusters better match the structures of neutral clusters. Isotopic mixtures of Ne atoms would make the experimental distinction between protonated and pure Ne clusters more difficult (this is not a problem for the nearly isotopically pure Ar), but we are nonetheless currently performing measurements on these systems with our setup.

 We have shown that protonated argon clusters show very different characteristics than pure, cationic argon clusters. The protonated clusters display magic sizes that perfectly match the magic cluster series reported by Harris \emph{et al.}\ \cite{Harris:1984aa,Harris:1986aa} for pure Ar$_n^+$ clusters, indicating that their measurements may have contained an unresolved contribution from protonated clusters, likely originating from some impurity (e.g.\ water) in their setup. This could thus solve the long standing disagreement between different studies on argon clusters regarding the nature and origin of abundance anomalies in the mass spectra \cite{Harris:1984aa,Harris:1986aa,Milne:1967aa,Ding:1983aa,Scheier:1987aa,Levinger:1988aa,Miehle:1989ab,Ferreira-da-Silva:2009aa}. It also highlights the dramatic differences small impurities can play in the formation and characteristics of clusters and small nanoparticles, similar to what has been observed regarding the electronic properties of small carbon cluster anions and their hydrides \cite{Ito:2014aa}, and the role that hydrogen plays in stabilizing metal clusters \cite{Kiran:2007aa}.
 
This work was supported by the Austrian Science Fund FWF (projects P26635 and W1259) and the Swedish Research Council (Contract No.\ 2016-06625). The computational results presented have been achieved (in part) using the HPC infrastructure LEO of the University of Innsbruck.

\bibliography{/Users/Michael/Dropbox/Documents/Bibtex/Library.bib}

\begin{thebibliography}{38}%
\makeatletter
\providecommand \@ifxundefined [1]{%
 \@ifx{#1\undefined}
}%
\providecommand \@ifnum [1]{%
 \ifnum #1\expandafter \@firstoftwo
 \else \expandafter \@secondoftwo
 \fi
}%
\providecommand \@ifx [1]{%
 \ifx #1\expandafter \@firstoftwo
 \else \expandafter \@secondoftwo
 \fi
}%
\providecommand \natexlab [1]{#1}%
\providecommand \enquote  [1]{``#1''}%
\providecommand \bibnamefont  [1]{#1}%
\providecommand \bibfnamefont [1]{#1}%
\providecommand \citenamefont [1]{#1}%
\providecommand \href@noop [0]{\@secondoftwo}%
\providecommand \href [0]{\begingroup \@sanitize@url \@href}%
\providecommand \@href[1]{\@@startlink{#1}\@@href}%
\providecommand \@@href[1]{\endgroup#1\@@endlink}%
\providecommand \@sanitize@url [0]{\catcode `\\12\catcode `\$12\catcode
  `\&12\catcode `\#12\catcode `\^12\catcode `\_12\catcode `\%12\relax}%
\providecommand \@@startlink[1]{}%
\providecommand \@@endlink[0]{}%
\providecommand \url  [0]{\begingroup\@sanitize@url \@url }%
\providecommand \@url [1]{\endgroup\@href {#1}{\urlprefix }}%
\providecommand \urlprefix  [0]{URL }%
\providecommand \Eprint [0]{\href }%
\providecommand \doibase [0]{http://dx.doi.org/}%
\providecommand \selectlanguage [0]{\@gobble}%
\providecommand \bibinfo  [0]{\@secondoftwo}%
\providecommand \bibfield  [0]{\@secondoftwo}%
\providecommand \translation [1]{[#1]}%
\providecommand \BibitemOpen [0]{}%
\providecommand \bibitemStop [0]{}%
\providecommand \bibitemNoStop [0]{.\EOS\space}%
\providecommand \EOS [0]{\spacefactor3000\relax}%
\providecommand \BibitemShut  [1]{\csname bibitem#1\endcsname}%
\let\auto@bib@innerbib\@empty
\bibitem [{\citenamefont {Echt}\ \emph {et~al.}(1981)\citenamefont {Echt},
  \citenamefont {Sattler},\ and\ \citenamefont {Recknagel}}]{Echt:1981aa}%
  \BibitemOpen
  \bibfield  {author} {\bibinfo {author} {\bibfnamefont {O.}~\bibnamefont
  {Echt}}, \bibinfo {author} {\bibfnamefont {K.}~\bibnamefont {Sattler}}, \
  and\ \bibinfo {author} {\bibfnamefont {E.}~\bibnamefont {Recknagel}},\ }\href
  {https://link.aps.org/doi/10.1103/PhysRevLett.47.1121} {\bibfield  {journal}
  {\bibinfo  {journal} {Physical Review Letters}\ }\textbf {\bibinfo {volume}
  {47}},\ \bibinfo {pages} {1121} (\bibinfo {year} {1981})}\BibitemShut
  {NoStop}%
\bibitem [{\citenamefont {Ikeshoji}\ \emph {et~al.}(1996)\citenamefont
  {Ikeshoji}, \citenamefont {Hafskjold}, \citenamefont {Hashi},\ and\
  \citenamefont {Kawazoe}}]{Ikeshoji:1996aa}%
  \BibitemOpen
  \bibfield  {author} {\bibinfo {author} {\bibfnamefont {T.}~\bibnamefont
  {Ikeshoji}}, \bibinfo {author} {\bibfnamefont {B.}~\bibnamefont {Hafskjold}},
  \bibinfo {author} {\bibfnamefont {Y.}~\bibnamefont {Hashi}}, \ and\ \bibinfo
  {author} {\bibfnamefont {Y.}~\bibnamefont {Kawazoe}},\ }\href
  {https://link.aps.org/doi/10.1103/PhysRevLett.76.1792} {\bibfield  {journal}
  {\bibinfo  {journal} {Physical Review Letters}\ }\textbf {\bibinfo {volume}
  {76}},\ \bibinfo {pages} {1792} (\bibinfo {year} {1996})}\BibitemShut
  {NoStop}%
\bibitem [{\citenamefont {Rey}\ \emph {et~al.}(1992)\citenamefont {Rey},
  \citenamefont {Gallego}, \citenamefont {I{\~n}iguez},\ and\ \citenamefont
  {Alonso}}]{Rey:1992aa}%
  \BibitemOpen
  \bibfield  {author} {\bibinfo {author} {\bibfnamefont {C.}~\bibnamefont
  {Rey}}, \bibinfo {author} {\bibfnamefont {L.~J.}\ \bibnamefont {Gallego}},
  \bibinfo {author} {\bibfnamefont {M.~P.}\ \bibnamefont {I{\~n}iguez}}, \ and\
  \bibinfo {author} {\bibfnamefont {J.~A.}\ \bibnamefont {Alonso}},\ }\href
  {\doibase https://doi.org/10.1016/0921-4526(92)90626-4} {\bibfield  {journal}
  {\bibinfo  {journal} {Physica B: Condensed Matter}\ }\textbf {\bibinfo
  {volume} {179}},\ \bibinfo {pages} {273} (\bibinfo {year}
  {1992})}\BibitemShut {NoStop}%
\bibitem [{\citenamefont {Leary}\ and\ \citenamefont
  {Doye}(1999)}]{Leary:1999aa}%
  \BibitemOpen
  \bibfield  {author} {\bibinfo {author} {\bibfnamefont {R.~H.}\ \bibnamefont
  {Leary}}\ and\ \bibinfo {author} {\bibfnamefont {J.~P.~K.}\ \bibnamefont
  {Doye}},\ }\href {\doibase 10.1103/PhysRevE.60.R6320} {\bibfield  {journal}
  {\bibinfo  {journal} {Physical Review E}\ }\textbf {\bibinfo {volume} {60}},\
  \bibinfo {pages} {R6320} (\bibinfo {year} {1999})}\BibitemShut {NoStop}%
\bibitem [{\citenamefont {Wales}\ and\ \citenamefont
  {Doye}(1997)}]{Wales:1997aa}%
  \BibitemOpen
  \bibfield  {author} {\bibinfo {author} {\bibfnamefont {D.~J.}\ \bibnamefont
  {Wales}}\ and\ \bibinfo {author} {\bibfnamefont {J.~P.~K.}\ \bibnamefont
  {Doye}},\ }\href {\doibase 10.1021/jp970984n} {\bibfield  {journal} {\bibinfo
   {journal} {The Journal of Physical Chemistry A}\ }\textbf {\bibinfo {volume}
  {101}},\ \bibinfo {pages} {5111} (\bibinfo {year} {1997})}\BibitemShut
  {NoStop}%
\bibitem [{\citenamefont {Xiang}\ \emph
  {et~al.}(2004{\natexlab{a}})\citenamefont {Xiang}, \citenamefont {Jiang},
  \citenamefont {Cai},\ and\ \citenamefont {Shao}}]{Xiang:2004ab}%
  \BibitemOpen
  \bibfield  {author} {\bibinfo {author} {\bibfnamefont {Y.}~\bibnamefont
  {Xiang}}, \bibinfo {author} {\bibfnamefont {H.}~\bibnamefont {Jiang}},
  \bibinfo {author} {\bibfnamefont {W.}~\bibnamefont {Cai}}, \ and\ \bibinfo
  {author} {\bibfnamefont {X.}~\bibnamefont {Shao}},\ }\href {\doibase
  10.1021/jp037780t} {\bibfield  {journal} {\bibinfo  {journal} {The Journal of
  Physical Chemistry A}\ }\textbf {\bibinfo {volume} {108}},\ \bibinfo {pages}
  {3586} (\bibinfo {year} {2004}{\natexlab{a}})}\BibitemShut {NoStop}%
\bibitem [{\citenamefont {Xiang}\ \emph
  {et~al.}(2004{\natexlab{b}})\citenamefont {Xiang}, \citenamefont {Cheng},
  \citenamefont {Cai},\ and\ \citenamefont {Shao}}]{Xiang:2004aa}%
  \BibitemOpen
  \bibfield  {author} {\bibinfo {author} {\bibfnamefont {Y.}~\bibnamefont
  {Xiang}}, \bibinfo {author} {\bibfnamefont {L.}~\bibnamefont {Cheng}},
  \bibinfo {author} {\bibfnamefont {W.}~\bibnamefont {Cai}}, \ and\ \bibinfo
  {author} {\bibfnamefont {X.}~\bibnamefont {Shao}},\ }\href {\doibase
  10.1021/jp047807o} {\bibfield  {journal} {\bibinfo  {journal} {The Journal of
  Physical Chemistry A}\ }\textbf {\bibinfo {volume} {108}},\ \bibinfo {pages}
  {9516} (\bibinfo {year} {2004}{\natexlab{b}})}\BibitemShut {NoStop}%
\bibitem [{\citenamefont {Br\"uhl}\ \emph {et~al.}(2004)\citenamefont
  {Br\"uhl}, \citenamefont {Guardiola}, \citenamefont {Kalinin}, \citenamefont
  {Kornilov}, \citenamefont {Navarro}, \citenamefont {Savas},\ and\
  \citenamefont {Toennies}}]{Bruhl:2004aa}%
  \BibitemOpen
  \bibfield  {author} {\bibinfo {author} {\bibfnamefont {R.}~\bibnamefont
  {Br\"uhl}}, \bibinfo {author} {\bibfnamefont {R.}~\bibnamefont {Guardiola}},
  \bibinfo {author} {\bibfnamefont {A.}~\bibnamefont {Kalinin}}, \bibinfo
  {author} {\bibfnamefont {O.}~\bibnamefont {Kornilov}}, \bibinfo {author}
  {\bibfnamefont {J.}~\bibnamefont {Navarro}}, \bibinfo {author} {\bibfnamefont
  {T.}~\bibnamefont {Savas}}, \ and\ \bibinfo {author} {\bibfnamefont {J.~P.}\
  \bibnamefont {Toennies}},\ }\href {\doibase 10.1103/PhysRevLett.92.185301}
  {\bibfield  {journal} {\bibinfo  {journal} {Physical Review Letters}\
  }\textbf {\bibinfo {volume} {92}},\ \bibinfo {pages} {185301} (\bibinfo
  {year} {2004})}\BibitemShut {NoStop}%
\bibitem [{\citenamefont {Stephens}\ and\ \citenamefont
  {King}(1983)}]{Stephens:1983aa}%
  \BibitemOpen
  \bibfield  {author} {\bibinfo {author} {\bibfnamefont {P.~W.}\ \bibnamefont
  {Stephens}}\ and\ \bibinfo {author} {\bibfnamefont {J.~G.}\ \bibnamefont
  {King}},\ }\href {\doibase 10.1103/PhysRevLett.51.1538} {\bibfield  {journal}
  {\bibinfo  {journal} {Physical Review Letters}\ }\textbf {\bibinfo {volume}
  {51}},\ \bibinfo {pages} {1538} (\bibinfo {year} {1983})}\BibitemShut
  {NoStop}%
\bibitem [{\citenamefont {Buchenau}\ \emph {et~al.}(1990)\citenamefont
  {Buchenau}, \citenamefont {Knuth}, \citenamefont {Northby}, \citenamefont
  {Toennies},\ and\ \citenamefont {Winkler}}]{Buchenau:1990aa}%
  \BibitemOpen
  \bibfield  {author} {\bibinfo {author} {\bibfnamefont {H.}~\bibnamefont
  {Buchenau}}, \bibinfo {author} {\bibfnamefont {E.~L.}\ \bibnamefont {Knuth}},
  \bibinfo {author} {\bibfnamefont {J.}~\bibnamefont {Northby}}, \bibinfo
  {author} {\bibfnamefont {J.~P.}\ \bibnamefont {Toennies}}, \ and\ \bibinfo
  {author} {\bibfnamefont {C.}~\bibnamefont {Winkler}},\ }\href {\doibase
  10.1063/1.458275} {\bibfield  {journal} {\bibinfo  {journal} {The Journal of
  Chemical Physics}\ }\textbf {\bibinfo {volume} {92}},\ \bibinfo {pages}
  {6875} (\bibinfo {year} {1990})}\BibitemShut {NoStop}%
\bibitem [{\citenamefont {Harris}\ \emph {et~al.}(1984)\citenamefont {Harris},
  \citenamefont {Kidwell},\ and\ \citenamefont {Northby}}]{Harris:1984aa}%
  \BibitemOpen
  \bibfield  {author} {\bibinfo {author} {\bibfnamefont {I.~A.}\ \bibnamefont
  {Harris}}, \bibinfo {author} {\bibfnamefont {R.~S.}\ \bibnamefont {Kidwell}},
  \ and\ \bibinfo {author} {\bibfnamefont {J.~A.}\ \bibnamefont {Northby}},\
  }\href {https://link.aps.org/doi/10.1103/PhysRevLett.53.2390} {\bibfield
  {journal} {\bibinfo  {journal} {Physical Review Letters}\ }\textbf {\bibinfo
  {volume} {53}},\ \bibinfo {pages} {2390} (\bibinfo {year}
  {1984})}\BibitemShut {NoStop}%
\bibitem [{\citenamefont {Harris}\ \emph {et~al.}(1986)\citenamefont {Harris},
  \citenamefont {Norman}, \citenamefont {Mulkern},\ and\ \citenamefont
  {Northby}}]{Harris:1986aa}%
  \BibitemOpen
  \bibfield  {author} {\bibinfo {author} {\bibfnamefont {I.~A.}\ \bibnamefont
  {Harris}}, \bibinfo {author} {\bibfnamefont {K.~A.}\ \bibnamefont {Norman}},
  \bibinfo {author} {\bibfnamefont {R.~V.}\ \bibnamefont {Mulkern}}, \ and\
  \bibinfo {author} {\bibfnamefont {J.~A.}\ \bibnamefont {Northby}},\ }\href
  {\doibase https://doi.org/10.1016/0009-2614(86)80476-6} {\bibfield  {journal}
  {\bibinfo  {journal} {Chemical Physics Letters}\ }\textbf {\bibinfo {volume}
  {130}},\ \bibinfo {pages} {316} (\bibinfo {year} {1986})}\BibitemShut
  {NoStop}%
\bibitem [{\citenamefont {M{\"a}rk}\ and\ \citenamefont
  {Scheier}(1987)}]{Mark:1987aa}%
  \BibitemOpen
  \bibfield  {author} {\bibinfo {author} {\bibfnamefont {T.~D.}\ \bibnamefont
  {M{\"a}rk}}\ and\ \bibinfo {author} {\bibfnamefont {P.}~\bibnamefont
  {Scheier}},\ }\href {\doibase https://doi.org/10.1016/0009-2614(87)80213-0}
  {\bibfield  {journal} {\bibinfo  {journal} {Chemical Physics Letters}\
  }\textbf {\bibinfo {volume} {137}},\ \bibinfo {pages} {245} (\bibinfo {year}
  {1987})}\BibitemShut {NoStop}%
\bibitem [{\citenamefont {Lezius}\ \emph {et~al.}(1989)\citenamefont {Lezius},
  \citenamefont {Scheier}, \citenamefont {Stamatovic},\ and\ \citenamefont
  {M{\"a}rk}}]{Lezius:1989aa}%
  \BibitemOpen
  \bibfield  {author} {\bibinfo {author} {\bibfnamefont {M.}~\bibnamefont
  {Lezius}}, \bibinfo {author} {\bibfnamefont {P.}~\bibnamefont {Scheier}},
  \bibinfo {author} {\bibfnamefont {A.}~\bibnamefont {Stamatovic}}, \ and\
  \bibinfo {author} {\bibfnamefont {T.~D.}\ \bibnamefont {M{\"a}rk}},\ }\href
  {\doibase 10.1063/1.456898} {\bibfield  {journal} {\bibinfo  {journal} {The
  Journal of Chemical Physics}\ }\textbf {\bibinfo {volume} {91}},\ \bibinfo
  {pages} {3240} (\bibinfo {year} {1989})}\BibitemShut {NoStop}%
\bibitem [{\citenamefont {Miehle}\ \emph {et~al.}(1989)\citenamefont {Miehle},
  \citenamefont {Kandler}, \citenamefont {Leisner},\ and\ \citenamefont
  {Echt}}]{Miehle:1989ab}%
  \BibitemOpen
  \bibfield  {author} {\bibinfo {author} {\bibfnamefont {W.}~\bibnamefont
  {Miehle}}, \bibinfo {author} {\bibfnamefont {O.}~\bibnamefont {Kandler}},
  \bibinfo {author} {\bibfnamefont {T.}~\bibnamefont {Leisner}}, \ and\
  \bibinfo {author} {\bibfnamefont {O.}~\bibnamefont {Echt}},\ }\href {\doibase
  10.1063/1.457464} {\bibfield  {journal} {\bibinfo  {journal} {The Journal of
  Chemical Physics}\ }\textbf {\bibinfo {volume} {91}},\ \bibinfo {pages}
  {5940} (\bibinfo {year} {1989})}\BibitemShut {NoStop}%
\bibitem [{\citenamefont {Milne}\ and\ \citenamefont
  {Greene}(1967)}]{Milne:1967aa}%
  \BibitemOpen
  \bibfield  {author} {\bibinfo {author} {\bibfnamefont {T.~A.}\ \bibnamefont
  {Milne}}\ and\ \bibinfo {author} {\bibfnamefont {F.~T.}\ \bibnamefont
  {Greene}},\ }\href {\doibase 10.1063/1.1701582} {\bibfield  {journal}
  {\bibinfo  {journal} {The Journal of Chemical Physics}\ }\textbf {\bibinfo
  {volume} {47}},\ \bibinfo {pages} {4095} (\bibinfo {year}
  {1967})}\BibitemShut {NoStop}%
\bibitem [{\citenamefont {Ding}\ and\ \citenamefont
  {Hesslich}(1983)}]{Ding:1983aa}%
  \BibitemOpen
  \bibfield  {author} {\bibinfo {author} {\bibfnamefont {A.}~\bibnamefont
  {Ding}}\ and\ \bibinfo {author} {\bibfnamefont {J.}~\bibnamefont
  {Hesslich}},\ }\href {\doibase https://doi.org/10.1016/0009-2614(83)87209-1}
  {\bibfield  {journal} {\bibinfo  {journal} {Chemical Physics Letters}\
  }\textbf {\bibinfo {volume} {94}},\ \bibinfo {pages} {54} (\bibinfo {year}
  {1983})}\BibitemShut {NoStop}%
\bibitem [{\citenamefont {Scheier}\ and\ \citenamefont
  {M{\"a}rk}(1987)}]{Scheier:1987aa}%
  \BibitemOpen
  \bibfield  {author} {\bibinfo {author} {\bibfnamefont {P.}~\bibnamefont
  {Scheier}}\ and\ \bibinfo {author} {\bibfnamefont {T.~D.}\ \bibnamefont
  {M{\"a}rk}},\ }\href {\doibase https://doi.org/10.1016/0168-1176(87)80031-9}
  {\bibfield  {journal} {\bibinfo  {journal} {International Journal of Mass
  Spectrometry and Ion Processes}\ }\textbf {\bibinfo {volume} {76}},\ \bibinfo
  {pages} {R11} (\bibinfo {year} {1987})}\BibitemShut {NoStop}%
\bibitem [{\citenamefont {Levinger}\ \emph {et~al.}(1988)\citenamefont
  {Levinger}, \citenamefont {Ray}, \citenamefont {Alexander},\ and\
  \citenamefont {Lineberger}}]{Levinger:1988aa}%
  \BibitemOpen
  \bibfield  {author} {\bibinfo {author} {\bibfnamefont {N.~E.}\ \bibnamefont
  {Levinger}}, \bibinfo {author} {\bibfnamefont {D.}~\bibnamefont {Ray}},
  \bibinfo {author} {\bibfnamefont {M.~L.}\ \bibnamefont {Alexander}}, \ and\
  \bibinfo {author} {\bibfnamefont {W.~C.}\ \bibnamefont {Lineberger}},\ }\href
  {\doibase 10.1063/1.455572} {\bibfield  {journal} {\bibinfo  {journal} {The
  Journal of Chemical Physics}\ }\textbf {\bibinfo {volume} {89}},\ \bibinfo
  {pages} {5654} (\bibinfo {year} {1988})}\BibitemShut {NoStop}%
\bibitem [{\citenamefont {Ferreira~da Silva}\ \emph {et~al.}(2009)\citenamefont
  {Ferreira~da Silva}, \citenamefont {Bartl}, \citenamefont {Denifl},
  \citenamefont {Echt}, \citenamefont {M{\"a}rk},\ and\ \citenamefont
  {Scheier}}]{Ferreira-da-Silva:2009aa}%
  \BibitemOpen
  \bibfield  {author} {\bibinfo {author} {\bibfnamefont {F.}~\bibnamefont
  {Ferreira~da Silva}}, \bibinfo {author} {\bibfnamefont {P.}~\bibnamefont
  {Bartl}}, \bibinfo {author} {\bibfnamefont {S.}~\bibnamefont {Denifl}},
  \bibinfo {author} {\bibfnamefont {O.}~\bibnamefont {Echt}}, \bibinfo {author}
  {\bibfnamefont {T.~D.}\ \bibnamefont {M{\"a}rk}}, \ and\ \bibinfo {author}
  {\bibfnamefont {P.}~\bibnamefont {Scheier}},\ }\href {\doibase
  10.1039/B913175B} {\bibfield  {journal} {\bibinfo  {journal} {Physical
  Chemistry Chemical Physics}\ }\textbf {\bibinfo {volume} {11}},\ \bibinfo
  {pages} {9791} (\bibinfo {year} {2009})}\BibitemShut {NoStop}%
\bibitem [{\citenamefont {Vafayi}\ and\ \citenamefont
  {Esfarjani}(2015)}]{Vafayi:2015aa}%
  \BibitemOpen
  \bibfield  {author} {\bibinfo {author} {\bibfnamefont {K.}~\bibnamefont
  {Vafayi}}\ and\ \bibinfo {author} {\bibfnamefont {K.}~\bibnamefont
  {Esfarjani}},\ }\href {\doibase 10.1007/s10876-014-0832-z} {\bibfield
  {journal} {\bibinfo  {journal} {Journal of Cluster Science}\ }\textbf
  {\bibinfo {volume} {26}},\ \bibinfo {pages} {473} (\bibinfo {year}
  {2015})}\BibitemShut {NoStop}%
\bibitem [{\citenamefont {Sch{\"o}bel}\ \emph
  {et~al.}(2011{\natexlab{a}})\citenamefont {Sch{\"o}bel}, \citenamefont
  {Bartl}, \citenamefont {Leidlmair}, \citenamefont {Denifl}, \citenamefont
  {Echt}, \citenamefont {M{\"a}rk},\ and\ \citenamefont
  {Scheier}}]{Schobel:2011aa}%
  \BibitemOpen
  \bibfield  {author} {\bibinfo {author} {\bibfnamefont {H.}~\bibnamefont
  {Sch{\"o}bel}}, \bibinfo {author} {\bibfnamefont {P.}~\bibnamefont {Bartl}},
  \bibinfo {author} {\bibfnamefont {C.}~\bibnamefont {Leidlmair}}, \bibinfo
  {author} {\bibfnamefont {S.}~\bibnamefont {Denifl}}, \bibinfo {author}
  {\bibfnamefont {O.}~\bibnamefont {Echt}}, \bibinfo {author} {\bibfnamefont
  {T.~D.}\ \bibnamefont {M{\"a}rk}}, \ and\ \bibinfo {author} {\bibfnamefont
  {P.}~\bibnamefont {Scheier}},\ }\href {\doibase 10.1140/epjd/e2011-10619-1}
  {\bibfield  {journal} {\bibinfo  {journal} {The European Physical Journal D}\
  }\textbf {\bibinfo {volume} {63}},\ \bibinfo {pages} {209} (\bibinfo {year}
  {2011}{\natexlab{a}})}\BibitemShut {NoStop}%
\bibitem [{\citenamefont {Kurzthaler}\ \emph {et~al.}(2016)\citenamefont
  {Kurzthaler}, \citenamefont {Rasul}, \citenamefont {Kuhn}, \citenamefont
  {Lindinger}, \citenamefont {Scheier},\ and\ \citenamefont
  {Ellis}}]{Kurzthaler:2016aa}%
  \BibitemOpen
  \bibfield  {author} {\bibinfo {author} {\bibfnamefont {T.}~\bibnamefont
  {Kurzthaler}}, \bibinfo {author} {\bibfnamefont {B.}~\bibnamefont {Rasul}},
  \bibinfo {author} {\bibfnamefont {M.}~\bibnamefont {Kuhn}}, \bibinfo {author}
  {\bibfnamefont {A.}~\bibnamefont {Lindinger}}, \bibinfo {author}
  {\bibfnamefont {P.}~\bibnamefont {Scheier}}, \ and\ \bibinfo {author}
  {\bibfnamefont {A.~M.}\ \bibnamefont {Ellis}},\ }\href {\doibase
  10.1063/1.4960611} {\bibfield  {journal} {\bibinfo  {journal} {The Journal of
  Chemical Physics}\ }\textbf {\bibinfo {volume} {145}},\ \bibinfo {pages}
  {064305} (\bibinfo {year} {2016})}\BibitemShut {NoStop}%
\bibitem [{\citenamefont {Kuhn}\ \emph {et~al.}(2016)\citenamefont {Kuhn},
  \citenamefont {Renzler}, \citenamefont {Postler}, \citenamefont {Ralser},
  \citenamefont {Spieler}, \citenamefont {Simpson}, \citenamefont {Linnartz},
  \citenamefont {Tielens}, \citenamefont {Cami}, \citenamefont {Mauracher},
  \citenamefont {Wang}, \citenamefont {Alcam{\'\i}}, \citenamefont
  {Mart{\'\i}n}, \citenamefont {Beyer}, \citenamefont {Wester}, \citenamefont
  {Lindinger},\ and\ \citenamefont {Scheier}}]{Kuhn:2016aa}%
  \BibitemOpen
  \bibfield  {author} {\bibinfo {author} {\bibfnamefont {M.}~\bibnamefont
  {Kuhn}}, \bibinfo {author} {\bibfnamefont {M.}~\bibnamefont {Renzler}},
  \bibinfo {author} {\bibfnamefont {J.}~\bibnamefont {Postler}}, \bibinfo
  {author} {\bibfnamefont {S.}~\bibnamefont {Ralser}}, \bibinfo {author}
  {\bibfnamefont {S.}~\bibnamefont {Spieler}}, \bibinfo {author} {\bibfnamefont
  {M.}~\bibnamefont {Simpson}}, \bibinfo {author} {\bibfnamefont
  {H.}~\bibnamefont {Linnartz}}, \bibinfo {author} {\bibfnamefont {A.~G.
  G.~M.}\ \bibnamefont {Tielens}}, \bibinfo {author} {\bibfnamefont
  {J.}~\bibnamefont {Cami}}, \bibinfo {author} {\bibfnamefont {A.}~\bibnamefont
  {Mauracher}}, \bibinfo {author} {\bibfnamefont {Y.}~\bibnamefont {Wang}},
  \bibinfo {author} {\bibfnamefont {M.}~\bibnamefont {Alcam{\'\i}}}, \bibinfo
  {author} {\bibfnamefont {F.}~\bibnamefont {Mart{\'\i}n}}, \bibinfo {author}
  {\bibfnamefont {M.~K.}\ \bibnamefont {Beyer}}, \bibinfo {author}
  {\bibfnamefont {R.}~\bibnamefont {Wester}}, \bibinfo {author} {\bibfnamefont
  {A.}~\bibnamefont {Lindinger}}, \ and\ \bibinfo {author} {\bibfnamefont
  {P.}~\bibnamefont {Scheier}},\ }\href {\doibase 10.1038/ncomms13550}
  {\bibfield  {journal} {\bibinfo  {journal} {Nature Communications}\ }\textbf
  {\bibinfo {volume} {7}},\ \bibinfo {pages} {13550} (\bibinfo {year}
  {2016})}\BibitemShut {NoStop}%
\bibitem [{\citenamefont {Ralser}\ \emph {et~al.}(2015)\citenamefont {Ralser},
  \citenamefont {Postler}, \citenamefont {Harnisch}, \citenamefont {Ellis},\
  and\ \citenamefont {Scheier}}]{Ralser:2015aa}%
  \BibitemOpen
  \bibfield  {author} {\bibinfo {author} {\bibfnamefont {S.}~\bibnamefont
  {Ralser}}, \bibinfo {author} {\bibfnamefont {J.}~\bibnamefont {Postler}},
  \bibinfo {author} {\bibfnamefont {M.}~\bibnamefont {Harnisch}}, \bibinfo
  {author} {\bibfnamefont {A.~M.}\ \bibnamefont {Ellis}}, \ and\ \bibinfo
  {author} {\bibfnamefont {P.}~\bibnamefont {Scheier}},\ }\href {\doibase
  https://doi.org/10.1016/j.ijms.2015.01.004} {\bibfield  {journal} {\bibinfo
  {journal} {International Journal of Mass Spectrometry}\ }\textbf {\bibinfo
  {volume} {379}},\ \bibinfo {pages} {194} (\bibinfo {year}
  {2015})}\BibitemShut {NoStop}%
\bibitem [{\citenamefont {Sch{\"o}bel}\ \emph
  {et~al.}(2011{\natexlab{b}})\citenamefont {Sch{\"o}bel}, \citenamefont
  {Bartl}, \citenamefont {Leidlmair}, \citenamefont {Denifl}, \citenamefont
  {Echt}, \citenamefont {M{\"a}rk},\ and\ \citenamefont
  {Scheier}}]{Schobel:2011ab}%
  \BibitemOpen
  \bibfield  {author} {\bibinfo {author} {\bibfnamefont {H.}~\bibnamefont
  {Sch{\"o}bel}}, \bibinfo {author} {\bibfnamefont {P.}~\bibnamefont {Bartl}},
  \bibinfo {author} {\bibfnamefont {C.}~\bibnamefont {Leidlmair}}, \bibinfo
  {author} {\bibfnamefont {S.}~\bibnamefont {Denifl}}, \bibinfo {author}
  {\bibfnamefont {O.}~\bibnamefont {Echt}}, \bibinfo {author} {\bibfnamefont
  {T.~D.}\ \bibnamefont {M{\"a}rk}}, \ and\ \bibinfo {author} {\bibfnamefont
  {P.}~\bibnamefont {Scheier}},\ }\href {\doibase 10.1140/epjd/e2011-10619-1}
  {\bibfield  {journal} {\bibinfo  {journal} {The European Physical Journal D}\
  }\textbf {\bibinfo {volume} {63}},\ \bibinfo {pages} {209} (\bibinfo {year}
  {2011}{\natexlab{b}})}\BibitemShut {NoStop}%
\bibitem [{\citenamefont {Hvistendahl}\ \emph {et~al.}(1990)\citenamefont
  {Hvistendahl}, \citenamefont {Saastad},\ and\ \citenamefont
  {Uggerud}}]{Hvistendahl:1990aa}%
  \BibitemOpen
  \bibfield  {author} {\bibinfo {author} {\bibfnamefont {G.}~\bibnamefont
  {Hvistendahl}}, \bibinfo {author} {\bibfnamefont {O.~W.}\ \bibnamefont
  {Saastad}}, \ and\ \bibinfo {author} {\bibfnamefont {E.}~\bibnamefont
  {Uggerud}},\ }\href {\doibase https://doi.org/10.1016/0168-1176(90)85016-U}
  {\bibfield  {journal} {\bibinfo  {journal} {International Journal of Mass
  Spectrometry and Ion Processes}\ }\textbf {\bibinfo {volume} {98}},\ \bibinfo
  {pages} {167} (\bibinfo {year} {1990})}\BibitemShut {NoStop}%
\bibitem [{\citenamefont {Lezius}\ \emph {et~al.}(1992)\citenamefont {Lezius},
  \citenamefont {Scheier},\ and\ \citenamefont {M{\"a}rk}}]{Lezius:1992aa}%
  \BibitemOpen
  \bibfield  {author} {\bibinfo {author} {\bibfnamefont {M.}~\bibnamefont
  {Lezius}}, \bibinfo {author} {\bibfnamefont {P.}~\bibnamefont {Scheier}}, \
  and\ \bibinfo {author} {\bibfnamefont {T.~D.}\ \bibnamefont {M{\"a}rk}},\
  }\href {\doibase https://doi.org/10.1016/0009-2614(92)85939-8} {\bibfield
  {journal} {\bibinfo  {journal} {Chemical Physics Letters}\ }\textbf {\bibinfo
  {volume} {196}},\ \bibinfo {pages} {118} (\bibinfo {year}
  {1992})}\BibitemShut {NoStop}%
\bibitem [{\citenamefont {Giju}\ \emph {et~al.}(2002)\citenamefont {Giju},
  \citenamefont {Roszak},\ and\ \citenamefont {Leszczynski}}]{Giju:2002aa}%
  \BibitemOpen
  \bibfield  {author} {\bibinfo {author} {\bibfnamefont {K.~T.}\ \bibnamefont
  {Giju}}, \bibinfo {author} {\bibfnamefont {S.}~\bibnamefont {Roszak}}, \ and\
  \bibinfo {author} {\bibfnamefont {J.}~\bibnamefont {Leszczynski}},\ }\href
  {\doibase 10.1063/1.1485956} {\bibfield  {journal} {\bibinfo  {journal} {The
  Journal of Chemical Physics}\ }\textbf {\bibinfo {volume} {117}},\ \bibinfo
  {pages} {4803} (\bibinfo {year} {2002})}\BibitemShut {NoStop}%
\bibitem [{\citenamefont {McDonald}\ \emph {et~al.}(2016)\citenamefont
  {McDonald}, \citenamefont {Mauney}, \citenamefont {Leicht}, \citenamefont
  {Marks}, \citenamefont {Tan}, \citenamefont {Kuo},\ and\ \citenamefont
  {Duncan}}]{McDonald:2016aa}%
  \BibitemOpen
  \bibfield  {author} {\bibinfo {author} {\bibfnamefont {D.~C.}\ \bibnamefont
  {McDonald}}, \bibinfo {author} {\bibfnamefont {D.~T.}\ \bibnamefont
  {Mauney}}, \bibinfo {author} {\bibfnamefont {D.}~\bibnamefont {Leicht}},
  \bibinfo {author} {\bibfnamefont {J.~H.}\ \bibnamefont {Marks}}, \bibinfo
  {author} {\bibfnamefont {J.~A.}\ \bibnamefont {Tan}}, \bibinfo {author}
  {\bibfnamefont {J.~L.}\ \bibnamefont {Kuo}}, \ and\ \bibinfo {author}
  {\bibfnamefont {M.~A.}\ \bibnamefont {Duncan}},\ }\href {\doibase
  10.1063/1.4972581} {\bibfield  {journal} {\bibinfo  {journal} {The Journal of
  Chemical Physics}\ }\textbf {\bibinfo {volume} {145}},\ \bibinfo {pages}
  {231101} (\bibinfo {year} {2016})}\BibitemShut {NoStop}%
\bibitem [{\citenamefont {Bondybey}\ and\ \citenamefont
  {Pimentel}(1972)}]{Bondybey:1972aa}%
  \BibitemOpen
  \bibfield  {author} {\bibinfo {author} {\bibfnamefont {V.~E.}\ \bibnamefont
  {Bondybey}}\ and\ \bibinfo {author} {\bibfnamefont {G.~C.}\ \bibnamefont
  {Pimentel}},\ }\href {\doibase 10.1063/1.1677786} {\bibfield  {journal}
  {\bibinfo  {journal} {The Journal of Chemical Physics}\ }\textbf {\bibinfo
  {volume} {56}},\ \bibinfo {pages} {3832} (\bibinfo {year}
  {1972})}\BibitemShut {NoStop}%
\bibitem [{\citenamefont {Johns}(1984)}]{Johns:1984aa}%
  \BibitemOpen
  \bibfield  {author} {\bibinfo {author} {\bibfnamefont {J.~W.~C.}\
  \bibnamefont {Johns}},\ }\href {\doibase
  https://doi.org/10.1016/0022-2852(84)90087-0} {\bibfield  {journal} {\bibinfo
   {journal} {Journal of Molecular Spectroscopy}\ }\textbf {\bibinfo {volume}
  {106}},\ \bibinfo {pages} {124} (\bibinfo {year} {1984})}\BibitemShut
  {NoStop}%
\bibitem [{\citenamefont {Kunttu}\ and\ \citenamefont
  {Seetula}(1994)}]{Kunttu:1994aa}%
  \BibitemOpen
  \bibfield  {author} {\bibinfo {author} {\bibfnamefont {H.~M.}\ \bibnamefont
  {Kunttu}}\ and\ \bibinfo {author} {\bibfnamefont {J.~A.}\ \bibnamefont
  {Seetula}},\ }\href {\doibase https://doi.org/10.1016/0301-0104(94)00273-8}
  {\bibfield  {journal} {\bibinfo  {journal} {Chemical Physics}\ }\textbf
  {\bibinfo {volume} {189}},\ \bibinfo {pages} {273} (\bibinfo {year}
  {1994})}\BibitemShut {NoStop}%
\bibitem [{\citenamefont {Ritschel}\ \emph {et~al.}(2005)\citenamefont
  {Ritschel}, \citenamefont {Kuntz},\ and\ \citenamefont
  {Z{\"u}licke}}]{Ritschel:2005aa}%
  \BibitemOpen
  \bibfield  {author} {\bibinfo {author} {\bibfnamefont {T.}~\bibnamefont
  {Ritschel}}, \bibinfo {author} {\bibfnamefont {P.~J.}\ \bibnamefont {Kuntz}},
  \ and\ \bibinfo {author} {\bibfnamefont {L.}~\bibnamefont {Z{\"u}licke}},\
  }\href {\doibase 10.1140/epjd/e2005-00070-4} {\bibfield  {journal} {\bibinfo
  {journal} {The European Physical Journal D - Atomic, Molecular, Optical and
  Plasma Physics}\ }\textbf {\bibinfo {volume} {33}},\ \bibinfo {pages} {421}
  (\bibinfo {year} {2005})}\BibitemShut {NoStop}%
\bibitem [{\citenamefont {Frisch}\ \emph {et~al.}(2016)\citenamefont {Frisch}
  \emph {et~al.}}]{Frisch:2016aa_short}%
  \BibitemOpen
  \bibfield  {author} {\bibinfo {author} {\bibfnamefont {M.~J.}\ \bibnamefont
  {Frisch}} \emph {et~al.},\ }\href@noop {} {\enquote {\bibinfo {title}
  {Gaussian 16 rev. a.03},}\ } (\bibinfo {year} {2016})\BibitemShut {NoStop}%
\bibitem [{\citenamefont {Ikegami}\ \emph {et~al.}(1993)\citenamefont
  {Ikegami}, \citenamefont {Kondow},\ and\ \citenamefont
  {Iwata}}]{Ikegami:1993aa}%
  \BibitemOpen
  \bibfield  {author} {\bibinfo {author} {\bibfnamefont {T.}~\bibnamefont
  {Ikegami}}, \bibinfo {author} {\bibfnamefont {T.}~\bibnamefont {Kondow}}, \
  and\ \bibinfo {author} {\bibfnamefont {S.}~\bibnamefont {Iwata}},\ }\href
  {\doibase 10.1063/1.464130} {\bibfield  {journal} {\bibinfo  {journal} {The
  Journal of Chemical Physics}\ }\textbf {\bibinfo {volume} {98}},\ \bibinfo
  {pages} {3038} (\bibinfo {year} {1993})}\BibitemShut {NoStop}%
\bibitem [{\citenamefont {Ito}\ \emph {et~al.}(2014)\citenamefont {Ito},
  \citenamefont {Furukawa}, \citenamefont {Tanuma}, \citenamefont {Matsumoto},
  \citenamefont {Shiromaru}, \citenamefont {Majima}, \citenamefont {Goto},
  \citenamefont {Azuma},\ and\ \citenamefont {Hansen}}]{Ito:2014aa}%
  \BibitemOpen
  \bibfield  {author} {\bibinfo {author} {\bibfnamefont {G.}~\bibnamefont
  {Ito}}, \bibinfo {author} {\bibfnamefont {T.}~\bibnamefont {Furukawa}},
  \bibinfo {author} {\bibfnamefont {H.}~\bibnamefont {Tanuma}}, \bibinfo
  {author} {\bibfnamefont {J.}~\bibnamefont {Matsumoto}}, \bibinfo {author}
  {\bibfnamefont {H.}~\bibnamefont {Shiromaru}}, \bibinfo {author}
  {\bibfnamefont {T.}~\bibnamefont {Majima}}, \bibinfo {author} {\bibfnamefont
  {M.}~\bibnamefont {Goto}}, \bibinfo {author} {\bibfnamefont {T.}~\bibnamefont
  {Azuma}}, \ and\ \bibinfo {author} {\bibfnamefont {K.}~\bibnamefont
  {Hansen}},\ }\href {\doibase 10.1103/PhysRevLett.112.183001} {\bibfield
  {journal} {\bibinfo  {journal} {Physical Review Letters}\ }\textbf {\bibinfo
  {volume} {112}},\ \bibinfo {pages} {183001} (\bibinfo {year}
  {2014})}\BibitemShut {NoStop}%
\bibitem [{\citenamefont {Kiran}\ \emph {et~al.}(2007)\citenamefont {Kiran},
  \citenamefont {Jena}, \citenamefont {Li}, \citenamefont {Grubisic},
  \citenamefont {Stokes}, \citenamefont {Gantef{\"o}r}, \citenamefont {Bowen},
  \citenamefont {Burgert},\ and\ \citenamefont {Schn{\"o}ckel}}]{Kiran:2007aa}%
  \BibitemOpen
  \bibfield  {author} {\bibinfo {author} {\bibfnamefont {B.}~\bibnamefont
  {Kiran}}, \bibinfo {author} {\bibfnamefont {P.}~\bibnamefont {Jena}},
  \bibinfo {author} {\bibfnamefont {X.}~\bibnamefont {Li}}, \bibinfo {author}
  {\bibfnamefont {A.}~\bibnamefont {Grubisic}}, \bibinfo {author}
  {\bibfnamefont {S.~T.}\ \bibnamefont {Stokes}}, \bibinfo {author}
  {\bibfnamefont {G.~F.}\ \bibnamefont {Gantef{\"o}r}}, \bibinfo {author}
  {\bibfnamefont {K.~H.}\ \bibnamefont {Bowen}}, \bibinfo {author}
  {\bibfnamefont {R.}~\bibnamefont {Burgert}}, \ and\ \bibinfo {author}
  {\bibfnamefont {H.}~\bibnamefont {Schn{\"o}ckel}},\ }\href
  {https://link.aps.org/doi/10.1103/PhysRevLett.98.256802} {\bibfield
  {journal} {\bibinfo  {journal} {Physical Review Letters}\ }\textbf {\bibinfo
  {volume} {98}},\ \bibinfo {pages} {256802} (\bibinfo {year}
  {2007})}\BibitemShut {NoStop}%
\end{thebibliography}%
\end{document}